\documentclass{article}
\usepackage{spconf,amsmath,graphicx,hyperref}
\usepackage[ruled,linesnumbered]{algorithm2e}
\usepackage{multirow}
\usepackage{booktabs}
\usepackage{amssymb} 
\usepackage{subcaption}
\usepackage{caption}
\usepackage[skip=2pt]{caption}

\title{Investigating training objectives for flow matching-based speech enhancement}
\name{Liusha Yang\textsuperscript{1}, Ziru Ge\textsuperscript{1}, Gui Zhang\textsuperscript{1}, Junan Zhang\textsuperscript{2}, Zhizheng Wu\textsuperscript{2*}\thanks{*Corresponding author}
}
\address{\textsuperscript{1} College of Big Data and Internet, Shenzhen Technology Universiy, China \\
\textsuperscript{2}The Chinese University of Hong Kong, Shenzhen, China \\
\texttt{yangliusha@sztu.edu.cn}, ~~\texttt{wuzhizheng@cuhk.edu.cn}}

\begin{document}
\topmargin=0mm
%
\maketitle
\begin{abstract}
Speech enhancement (SE) aims to recover clean speech from noisy recordings. Although generative approaches such as score matching and the Schrödinger bridge have shown strong effectiveness, they are often computationally expensive. Flow matching offers a more efficient alternative by directly learning a velocity field that maps noise to data. In this work, we present a systematic study of flow matching for SE under three training objectives: velocity prediction, $\mathbf{x}_1$ prediction, and preconditioned $\mathbf{x}_1$ prediction. We analyze their impact on training dynamics and overall performance. Moreover, by introducing perceptual (PESQ) and signal-based (SI-SDR) objectives, we further enhance convergence efficiency and speech quality, yielding substantial improvements across evaluation metrics.
\end{abstract}
\section{Introduction}

Speech enhancement (SE)—the task of recovering clean signals from noisy recordings—has long been a core challenge in speech processing. Early statistical methods\,\cite{statistical2002} have been largely outperformed by deep neural networks (DNNs), which directly estimate clean waveforms or time–frequency masks\,\cite{Demucs2020,frcrn2022,CMGAN2024}. Despite their empirical success, these discriminative approaches do not explicitly capture the underlying generative structure of speech, potentially limiting their generalization across varied noise conditions.

In response, recent years have witnessed a growing interest in generative approaches for SE\,\cite{richter2023speech,SB2024,lee2025flowse}. Among them, score matching, Schrödinger bridge, and flow matching (FM) have emerged as particularly influential directions. Score matching and Schrödinger bridge are theoretically well-founded but remain computationally expensive due to their reliance on a large number of sampling steps or iterative optimization. In contrast, FM learns a vector field that directly transports noise into data, offering both stable training and efficient inference. These properties have motivated several concurrent efforts to explore FM for SE~\cite{lee2025flowse,wang2025flowse,korostik2025modifying}, which already suggest strong potential. 

The training objective in FM is typically formulated as regressing onto a target vector field along a probability path, a setting known as velocity prediction~\cite{lipman2024flow}. Within the family of affine conditional flows, the vector field can alternatively be parameterized as a function of the target data itself, referred to as ${\bf x}_1$ prediction in \cite{lipman2024flow}. Building on this, a preconditioned ${\bf x}_1$ prediction variant under the EDM framework~\cite{karras2022elucidating} improves both training stability and efficiency. Theoretically, following arguments in \cite{kingma2023understanding}, ${\bf x}_1$ prediction and its preconditioned variant are theoretically equivalent to velocity prediction but differ in their loss reweighting across time steps. However, their training behaviors and downstream performance—particularly in speech enhancement—remain insufficiently understood.


In this paper, we conduct a systematic study of FM for SE under different training objectives, examining their effects on training stability, convergence speed, and enhancement quality. Our results show that preconditioned ${\bf x}_1$ prediction performs comparably to velocity prediction but substantially outperforms vanilla ${\bf x}_1$ prediction across all metrics. Notably, it enables convergence at nearly twice the speed of velocity prediction, offering a favorable efficiency–accuracy trade-off. Furthermore, by incorporating task-specific objectives—perceptual evaluation of speech quality (PESQ) and scale-invariant signal-to-distortion ratio (SI-SDR) losses, we further improve both optimization efficiency and perceptual quality. Finally, comparing with other generative baselines, FM with preconditioned ${\bf x}_1$ prediction and task-specific objectives achieves the most balanced performance across metrics.

\vspace{-0.2cm}
\section{Flow matching for speech enhancement}
Flow Matching~\cite{lipman2024flow} defines a probability path $(p_t)_{0 \leq t \leq 1}$ that 
transforms a source distribution $p_0$ into a target 
distribution $p_1$, given only samples from $p_1$. A flow model is a continuous-time 
process ${\bf x}_t={\pmb\phi}_t({\bf x}_0)$, where ${\bf x}_0 \sim p_0$ and the flow ${\pmb\phi}_t$ is governed 
by a vector field ${\bf v}_t$ via the ordinary differential equation (ODE):
\vspace{-0.2cm}
\begin{equation}\nonumber
\frac{d}{dt}{\pmb\phi}_t({\bf x}_0)={\bf v}_t({\pmb\phi}_t({\bf x}_0)), \quad {\pmb\phi}_0({\bf x}_0)={\bf x}_0.
\vspace{-0.1cm}
\end{equation}
The induced density is
\vspace{-0.2cm}
\begin{equation}\nonumber
p_t({\bf x}_t)=p_0({\pmb\phi}_t^{-1}({\bf x}_t))\det\!\left[\frac{\partial {\pmb\phi}_t^{-1}({\bf x}_t)}{\partial {\bf x}_t}\right].
\vspace{-0.2cm}
\end{equation}

Let ${\bf y}$ and ${\bf x}_1$ denote the noisy speech signal and its clean counterpart. A flow matching-based speech enhancement model aims to construct a conditional probability path
\vspace{-0.1cm}
\begin{equation}\nonumber
p_t({\bf x}_t|{\bf x}_1, {\bf y}) = \mathcal{N}(\pmb{\mu}_t({\bf x}_1, {\bf y}), \sigma_t^2 {\bf I}),
\vspace{-0.1cm}
\end{equation}
that smoothly interpolates between the noisy input distribution $p_0$ and the clean target distribution $p_1$:
\vspace{-0.1cm}
\begin{align}\nonumber
p_0({\bf x}_0|{\bf x}_1, {\bf y}) = \mathcal{N}({\bf y}, \sigma_0^2{\bf I}), ~~
p_1({\bf x}_1|{\bf x}_1, {\bf y}) = \mathcal{N}({\bf x}_1, \sigma_1^2{\bf I}).
\end{align}
Here, we choose the conditional Optimal Transport flow, a particular instance belonging to the family of affine conditional flows, for its simplicity, faster training and better generation~\cite{lipman2024flow}, where $\pmb{\mu}_t$ and $\sigma_t$ change linearly in time, i.e.,
\vspace{-0.1cm}
\begin{align}\nonumber
\pmb{\mu}_t({\bf x}_1, {\bf y}) = t{\bf x}_1 + (1-t){\bf y}, \quad
\sigma_t = (1-t)\sigma_{\rm max}.
\vspace{-0.5cm}
\end{align}
The corresponding conditional vector field is derived as
\vspace{-0.2cm}
\begin{align}\label{velocity}
{\bf v}_t({\bf x}_t|{\bf x}_1, {\bf y}) = \frac{{\bf x}_t - {\bf x}_1}{1-t},
\vspace{-0.2cm}
\end{align}
which describes the instantaneous direction for transporting samples ${\bf x}_t$ toward the clean signal ${\bf x}_1$. By approximating this vector field with a neural network-based estimator ${\bf v}_\theta({\bf x}_t, {\bf y}, t)$, one can recover ${\bf x}_1$ by numerically solving the ODE
\vspace{-0.2cm}
\begin{equation}\label{ODE}
\frac{d}{dt} {\bf x}_t = {\bf v}_\theta({\bf x}_t, {\bf y}, t),
\vspace{-0.1cm}
\end{equation}
initialized from noisy samples ${\bf x}_0 \sim p_0({\bf x}_0|{\bf x}_1, {\bf y})$. 

The estimation of the vector field ${\bf v}_t$ can be formulated in multiple ways, depending on the chosen training objective. For instance, one may directly regress the velocity field~\cite{lee2025flowse}, predict the clean target ${\bf x}_1$~\cite{korostik2025modifying}, or adopt a preconditioned variant~\cite{karras2022elucidating}. Each formulation induces distinct optimization dynamics, influencing factors such as training stability, convergence rate, and generalization performance. In this work, we systematically investigate these variants to identify effective strategies for flow matching-based speech enhancement.

\section{Training objectives for flow matching-based speech enhancement}

We adopt the same data representation as in \cite{richter2023speech}, where clean speech ${\bf x}_1$ and noisy speech ${\bf y}$ are represented in the complex-valued STFT domain. To mitigate the heavy-tailed nature of the STFT coefficients $c$, an amplitude transformation is applied to each coefficient:
\vspace{-0.2cm}
\begin{align}\label{amp_copress}
\tilde{c} = \beta |c|^{\alpha} e^{i \angle(c)},
\vspace{-0.2cm}
\end{align}
where $\beta$ is a scaling factor and $\alpha$ is a compression factor. 
The conditional flow matching model is then trained to estimate the clean spectrograms from their noisy counterparts.

\subsection{The velocity prediction objective} 
The standard flow matching framework trains a neural network to approximate the conditional vector field ${\bf v}_t$ by minimizing the following mean-squared error:
\begin{equation}\nonumber
\mathcal{L}_{{\rm CFM}\text{-}{\bf v}} := 
\mathbb{E}_{t,\,{\bf x}_1,{\bf y}, \,p_t({\bf x}_t \mid {\bf x}_1, {\bf y})}
\bigl\| {\bf v}_\theta({\bf x}_t, {\bf y}, t) - {\bf v}_t({\bf x}_t \mid {\bf x}_1, {\bf y}) \bigr\|^2 .
\end{equation}
This objective, adopted in FlowSE \cite{lee2025flowse}, directly encourages the network to learn the transport dynamics that map noisy features toward clean ones. It is conceptually simple and leads to stable training in practice.

\subsection{The ${\bf x}_1$-prediction objective}
An alternative approach is to directly predict the clean speech ${\bf x}_1$, rather than the vector field. This strategy, referred to as ${\bf x}_1$ prediction, has also been adopted in \cite{korostik2025modifying} for speech enhancement. The rationale is that predicting structured speech signals may be easier for a neural network than predicting an abstract vector field. The training objective is formulated as
\begin{equation}\nonumber
\mathcal{L}_{{\rm CFM}\text{-}{\bf x}_1} := 
\mathbb{E}_{t,\,{\bf x}_1,{\bf y},\,p_t({\bf x}_t \mid {\bf x}_1, {\bf y})}
\bigl\| {\bf x}_\theta({\bf x}_t, {\bf y}, t) - {\bf x}_1 \bigr\|^2 .
\end{equation}

In this case, the vector field is estimated indirectly by substituting the network output ${\bf x}_\theta$ into the expression of the target in (\ref{velocity}), yielding
\vspace{-0.2cm}
\begin{align}\nonumber
\hat{\bf v}_t = \frac{{\bf x}_t- {\bf x}_\theta}{1-t}.
\vspace{-0.2cm}
\end{align}

During inference, the ODE in (\ref{ODE}) is numerically solved with $\hat{\bf v}_t$ computed from ${\bf x}_\theta$.

\subsection{The preconditioned ${\bf x}_1$-prediction objective}

Preconditioning refers to rescaling or transforming the network’s inputs and outputs at each time step to improve numerical stability and training efficiency. We adopt the preconditioning scheme of the EDM framework \cite{karras2022elucidating}, which is motivated by three principles: (i) maintaining unit variance for inputs, (ii) maintaining unit variance for targets, and (iii) ensuring uniform effective loss weighting across time.

Let $D_\theta$ denote the denoising model trained to minimize a weighted $L_2$ loss between the clean training data ${\bf x}_1$ and the model output:
\vspace{-0.2cm}
\begin{equation}\nonumber
\mathcal{L}_{{\rm CFM}\text{-}{\bf x}_1\text{-EDM}} := \mathbb{E}_{t,\,{\bf x}_1,{\bf y},\,p_t({\bf x}_t \mid {\bf x}_1, {\bf y})}
\left[\lambda(t)\bigl\| D_\theta({\bf x}_t, {\bf y}, t) - {\bf x}_1 \bigr\|^2\right].
\vspace{-0.2cm}
\end{equation}
The denoising model $D_\theta$ is parameterized as
\vspace{-0.15cm}
\begin{equation}\nonumber
D_\theta({\bf x}_t, {\bf y}, t) =
c_{\rm skip}(t)\,{\bf x}_t +
c_{\rm out}(t)\,F_\theta\!\left(c_{\rm in}(t){\bf x}_t,\, c_{\rm in}(t){\bf y},\, t\right),
\vspace{-0.15cm}
\end{equation}
where $F_\theta$ denotes the neural network backbone. The coefficients are given by 
\vspace{-0.2cm}
\small
\begin{align}\nonumber
c_{\rm skip}(t) &= \frac{\sigma_{\rm data}^2}{\sigma_{\rm data}^2+t^2\sigma_{\rm max}^2}, ~~
c_{\rm out}(t) = \frac{t\sigma_{\rm max}\sigma_{\rm data}}{\sqrt{\sigma_{\rm data}^2+t^2\sigma_{\rm max}^2}}, ~~\\\nonumber
c_{\rm in}(t)  &= \frac{1}{\sqrt{\sigma_{\rm data}^2+t^2\sigma_{\rm max}^2}}, 
~~\,\lambda(t) = \frac{t^2\sigma_{\rm max}^2+\sigma_{\rm data}^2}{t^2\sigma_{\rm max}^2\sigma_{\rm data}^2},
\vspace{-0.2cm}
\end{align}
\normalsize
where $\sigma_{\rm data}^2$ denotes the variance of the clean speech data distribution $p({\bf x}_1)$.

\subsection{Additional PESQ and SI-SDR loss}
In addition to the previously introduced training objectives, we incorporate two additional loss terms to further improve the perceptual quality of the enhanced speech. 

The first one is the PESQ loss, denoted as $L_{\rm PESQ}$. PESQ is a widely used objective metric in speech enhancement that correlates with human perception. It has been employed in training deep speech enhancement models either via adversarial learning \cite{fu2019metricgan,tsao2021metricgan+} or through differentiable implementations of the PESQ function \cite{richter2025investigating}. In this work, we adopt the differentiable PyTorch implementation of PESQ\footnote{\url{https://github.com/audiolabs/torch-pesq}}, 
and define the PESQ loss as
\vspace{-0.cm}
\begin{equation}\nonumber
\mathcal{L}_{\rm PESQ}(\underline{\hat{\bf x}}_1,\underline{\bf x}_1),
\vspace{-0.cm}
\end{equation}
where $\underline{\hat{\bf x}}_1 = \text{iSTFT}(\hat{\bf x}_1)$ and $\underline{\bf x}_1 = \text{iSTFT}({\bf x}_1)$ are the corresponding time-domain signals reconstructed by the inverse short-time Fourier transform (iSTFT). 

However, prior studies \cite{de2024pesqetarian} have shown that directly optimizing PESQ may lead to unnatural distortions in the enhanced speech, often yielding unreasonably high PESQ scores. To mitigate this effect, we additionally introduce the SI-SDR loss, denoted as $L_{\text{SI-SDR}}$, which is defined as
\vspace{-0.1cm}
\begin{equation}\nonumber
\mathcal{L}_{\text{SI-SDR}}(\underline{\hat{\bf x}}_1,\underline{\bf x}_1) = -10\log_{10}\frac{\|\omega \,\underline{\bf x}_1\|_2^2}
{\|\underline{\hat{\bf x}}_1 - \omega \underline{\bf x}_1\|_2^2},
\vspace{-0.1cm}
\end{equation}
with the optimal scale factor $\omega = \frac{\underline{\hat{\bf x}}_1^T \underline{\bf x}_1}
{\underline{\bf x}_1^T \underline{\bf x}_1}$.

The PESQ and SI-SDR losses can be combined with any of the previously defined training objectives, including $\mathcal{L}_{{\rm CFM}\text{-}{\bf v}}$, $\mathcal{L}_{{\rm CFM}\text{-}{\bf x}_1}$, and $\mathcal{L}_{{\rm CFM}\text{-}{\bf x}_1\text{-EDM}}$. 
When adopting the velocity prediction objective, the model does not directly output $\hat{\bf x}_1$. In this case, we first transform the estimated vector field into a signal estimate via
\vspace{-0.2cm}
\begin{equation}\nonumber
\hat{\bf x}_1 = {\bf x}_t + (1-t)\,{\bf v}_\theta({\bf x}_t, {\bf y}, t).
\vspace{-0.2cm}
\end{equation}
The overall objective then becomes
\vspace{-0.2cm}
\begin{equation}\nonumber
\mathcal{L}_{{\rm CFM}\text{-}{\bf v}\text{-PESQ-SISDR}} = \mathcal{L}_{{\rm CFM}\text{-}{\bf v}} + \alpha_p \mathcal{L}_{\rm PESQ} + \alpha_s \mathcal{L}_{\text{SI-SDR}},
\vspace{-0.2cm}
\end{equation}
where $\alpha_p$ and $\alpha_s$ are hyperparameters controlling the trade-off. 
For ${\bf x}_1$-based objectives, $\hat{\bf x}_1$ is explicitly predicted, so PESQ loss and SI-SDR loss can be directly applied.

\section{Experimental Setup}
\noindent\textbf{Data.}
Training and evaluation are based on the VoiceBank-DEMAND dataset \cite{botinhao2016investigating}, which provides paired clean and noisy 
recordings at 48\,kHz. All signals are resampled to 16\,kHz. Following \cite{richter2023speech}, we exclude speakers \texttt{p226} and \texttt{p287} from training and use them for validation.  

\noindent\textbf{Hyperparameters and training configuration.}
Audio signals are transformed into complex-valued STFT representations with a 510-point window, 128 hop size, and periodic Hann window, yielding $F=256$ frequency bins. We apply the amplitude transformation in (\ref{amp_copress}) with $\alpha=0.5$ and $\beta=0.15$, the same settings as in \cite{richter2023speech}. 

We adopt the Noise Conditional Score Network (NCSN++) as the backbone architecture, following the same configuration as \cite{richter2023speech, lay2023reducing, lee2025flowse}. Training is performed with the Adam optimizer (learning rate $10^{-4}$, batch size 32). An exponential moving average with decay rate 0.999 is applied to stabilize optimization. During training, the time parameter is sampled as $t \sim \mathcal{U}(t_\epsilon,1)$ with $t_\epsilon=0.03$, as in \cite{lee2025flowse}, to avoid instability at very small~$t$. $\sigma_{\rm max}$ and $\sigma_{\rm data}$ are determined empirically to $0.5$ and $0.1$.

\noindent\textbf{Objective metrics.}
We evaluate the performance of the systems in terms of PESQ, extended short-term objective intelligibility (ESTOI), SI-SNR, deep noise suppression MOS (DNSMOS) P.835 including SIG, BAK and OVRL, and word error rate (WER), which is computed using the Whisper large-v3 speech recognition model. 

\noindent\textbf{Baselines.}
We compare with diffusion-based SE models: SGMSE+ \cite{richter2023speech}, StoRM \cite{lemercier2023storm}, and SBVE \cite{richter2025investigating}.

\begin{table*}[t]
    \centering
    \small
    \caption{Comparison of different flow matching training objectives for speech enhancement. 
    Best results are highlighted in \textbf{bold}.}
    \vspace{-3pt}
    \begin{tabular}{cccccccc}
    \toprule
         Objective & PESQ  & ESTOI  & SI-SDR & SIG & BAK & OVRL & WER \\
    \midrule
         $\mathcal{L}_{{\rm CFM}\text{-}{\bf v}}$& 3.04$\pm$0.04 & \textbf{0.87$\pm$0.00} & \textbf{19.10$\pm$0.22} & \textbf{3.49$\pm$0.01} & \textbf{4.05$\pm$0.01} & \textbf{3.21$\pm$0.01} & 0.025$\pm$0.01 \\
         $\mathcal{L}_{{\rm CFM}\text{-}{\bf x}_1}$& 2.93$\pm$0.04 & 0.86$\pm$0.00 & 18.71$\pm$0.22 & 3.46$\pm$0.01 & 4.04$\pm$0.01 & 3.18$\pm$0.01 & 0.030$\pm$0.00 \\
         $\mathcal{L}_{{\rm CFM}\text{-}{\bf x}_1\text{-EDM}}$& \textbf{3.08$\pm$0.04} & \textbf{0.87$\pm$0.00} & 18.83$\pm$0.22 & \textbf{3.49$\pm$0.01} & 4.04$\pm$0.01 & 3.20$\pm$0.01 & \textbf{0.024$\pm$0.00} \\
    \bottomrule
    \end{tabular} 
    \label{tab:diffobj}
\end{table*}

\begin{table*}[ht!]
\centering
\small
\caption{Speech enhancement results with additional PESQ and SI-SDR losses. Best results are highlighted in \textbf{bold}.}
\renewcommand{\arraystretch}{0.8}
\setlength{\tabcolsep}{3pt} 
\begin{tabular}{c c c c c c c c c c}
\toprule
Objective & $\alpha_p$ & $\alpha_s$ & PESQ  & ESTOI  & SI-SDR & SIG & BAK & OVRL & WER \\
\midrule
\multicolumn{10}{c}{Additional $\mathcal{L}_{\rm PESQ}$} \\
\midrule
$\mathcal{L}_{{\rm CFM}\text{-}{\bf v}}$ & 5e-2 & 0 & 3.67$\pm$0.03 & 0.83$\pm$0.00 & 5.19$\pm$0.19 & 3.31$\pm$0.01 & 4.06$\pm$0.01 & 3.06$\pm$0.01 & 0.053$\pm$0.01 \\
$\mathcal{L}_{{\rm CFM}\text{-}{\bf x}_1}$ & 1e-3  & 0 & 3.64$\pm$0.03 & 0.84$\pm$0.00 & 6.71$\pm$0.19 & 3.37$\pm$0.01 & \textbf{4.08$\pm$0.01} & 3.12$\pm$0.01 & 0.037$\pm$0.01  \\
$\mathcal{L}_{{\rm CFM}\text{-}{\bf x}_1\text{-EDM}}$  & 1e-6  & 0 & \textbf{3.70$\pm$0.03} & 0.84$\pm$0.00 & 5.33$\pm$0.17 & 3.32$\pm$0.01 & 4.06$\pm$0.01 & {3.07$\pm$0.01} & 0.040$\pm$0.01  \\
\midrule
\multicolumn{10}{c}{Additional $\mathcal{L}_{\rm PESQ}$ and $\mathcal{L}_{\rm SISDR}$} \\
\midrule
$\mathcal{L}_{{\rm CFM}\text{-}{\bf v}}$ & 5e-2 & 5e-3 & 3.30$\pm$0.04 & \textbf{0.87$\pm$0.00} & 18.56$\pm$0.22 & \textbf{3.50$\pm$0.01} & \textbf{4.08$\pm$0.01} & \textbf{3.23$\pm$0.01} & \textbf{0.026$\pm$0.01} \\
$\mathcal{L}_{{\rm CFM}\text{-}{\bf x}_1}$ & 1e-3  & 1e-4 & 3.26$\pm$0.04 & \textbf{0.87$\pm$0.00} & \textbf{18.98$\pm$0.22} & \textbf{3.50$\pm$0.01} & 4.07$\pm$0.01 & \textbf{3.23$\pm$0.01} & 0.027$\pm$0.01  \\
$\mathcal{L}_{{\rm CFM}\text{-}{\bf x}_1\text{-EDM}}$  & 1e-6  & 1e-7 & {3.34$\pm$0.04} & \textbf{0.87$\pm$0.00} & 18.57$\pm$0.22 & \textbf{3.50$\pm$0.01} & \textbf{4.08$\pm$0.01} & \textbf{3.23$\pm$0.01} & 0.027$\pm$0.01  \\
\bottomrule
\end{tabular}
\label{tab:pesq_sisdr}
\end{table*}

\begin{table*}[ht!]
\centering
\small
\caption{Comparison with state-of-the-art generative speech enhancement baselines. Best results are highlighted in \textbf{bold}, and second best are \underline{underlined}. In case all methods achieve the same score on a metric, no highlighting is applied.}
\renewcommand{\arraystretch}{0.8}
\setlength{\tabcolsep}{3pt} 
\begin{tabular}{c c c c c c c c c c c}
\toprule
Method & $N$ & $\alpha_p$ & $\alpha_s$ & PESQ  & ESTOI  & SI-SDR & SIG & BAK & OVRL & WER \\
\midrule
SGMSE+ & 30 & 0  & 0 & 2.91$\pm$0.04 & 0.87$\pm$0.00 & 17.28$\pm$0.22 & 3.49$\pm$0.01 & 3.98$\pm$0.01 & 3.17$\pm$0.01 & 0.040$\pm$0.01  \\
StoRM & 50 & 0  & 0 & 2.87$\pm$0.04 & 0.87$\pm$0.00 & 18.48$\pm$0.22 & 3.49$\pm$0.01 & 4.03$\pm$0.01 & 3.20$\pm$0.01 & 0.037$\pm$0.01  \\
SBVE & 30 & 0 & 0 & 2.91$\pm$0.04 & 0.87$\pm$0.00 & \textbf{19.43$\pm$0.22} & 3.49$\pm$0.01 & 4.06$\pm$0.01 & \underline{3.22$\pm$0.01} & \underline{0.026$\pm$0.01} \\
FM w/ $\mathcal{L}_{{\rm CFM}\text{-}{\bf x}_1\text{-EDM}}$ & 5 & 0 & 0 & {3.08$\pm$0.04} & {0.87$\pm$0.00} & \underline{18.83$\pm$0.22} & {3.49$\pm$0.01} & 4.04$\pm$0.01 & 3.20$\pm$0.01 & \textbf{0.024$\pm$0.00} \\
\midrule
SBVE  & 30 & 5e-4  & 0 & \textbf{3.56$\pm$0.04} & 0.86$\pm$0.00 & 13.20$\pm$0.19 & 3.46$\pm$0.01 & \textbf{4.09$\pm$0.01} & 3.21$\pm$0.01 & 0.028$\pm$0.01  \\
FM w/ $\mathcal{L}_{{\rm CFM}\text{-}{\bf x}_1\text{-EDM}}$  & 5 & 1e-6  & 1e-7 & \underline{3.34$\pm$0.04} & 0.87$\pm$0.00 & {18.57$\pm$0.22} & \textbf{3.50$\pm$0.01} & \underline{4.08$\pm$0.01} & \textbf{3.23$\pm$0.01} & {0.027$\pm$0.01}  \\
\bottomrule
\end{tabular}
\label{tab:baselines}
\end{table*}
\vspace{-0.2cm}
\section{Results}
\subsection{Comparison of training objectives}\label{sec:exp_obj}
\begin{figure}[htbp]
    \centering  
    \begin{subfigure}[t]{0.5\linewidth} 
        \centering
        \includegraphics[width=\linewidth]{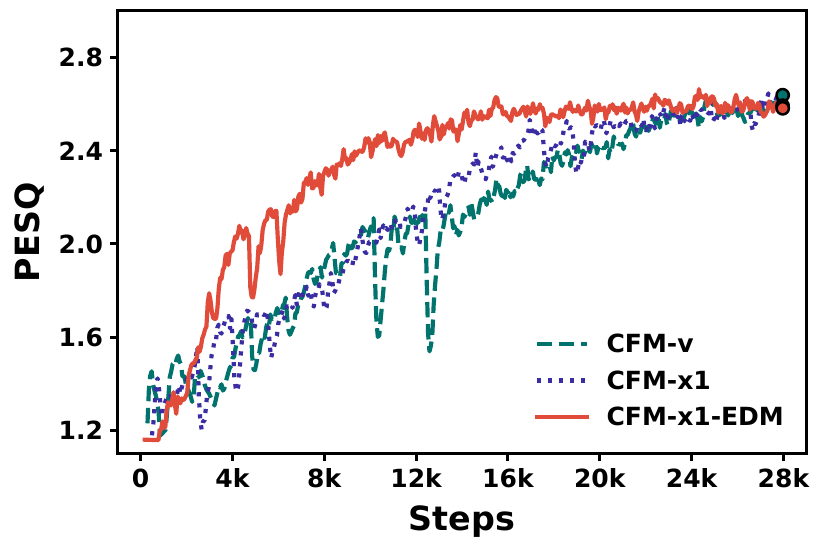}
        \caption{\footnotesize Objective-only loss}
        \label{fig:pesq_sisdr1}
    \end{subfigure}%
    \begin{subfigure}[t]{0.5\linewidth} 
        \centering
        \includegraphics[width=\linewidth]{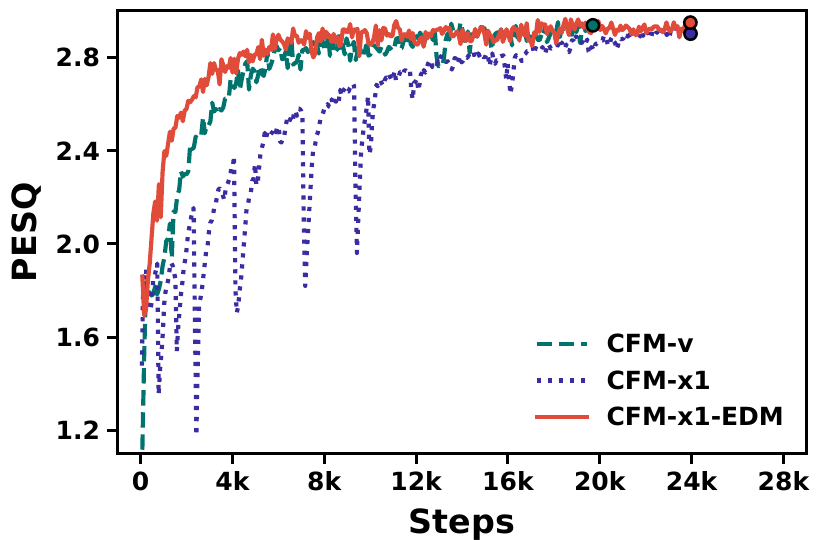}
        \caption{\footnotesize\mbox{Objective\,+\,PESQ\,+SI-SDR\,losses}}
        \label{fig:pesq_sisdr2}
    \end{subfigure}
    \caption{Performance in PESQ on the validation dataset over the training steps.}
\end{figure}

We first evaluate flow matching models trained with three different objectives: the velocity objective, the $\mathbf{x}_1$ objective, and the preconditioned $\mathbf{x}_1$ objective. For fair comparison, the number of reverse sampling steps is fixed to $N=5$ for all models. Results are reported as mean $\pm$ $95\%$ confidence interval across test samples. As summarized in Table\,\ref{tab:diffobj}, models trained with $\mathcal{L}_{{\rm CFM}\text{-}{\bf v}}$ and $\mathcal{L}_{{\rm CFM}\text{-}{\bf x}_1\text{-EDM}}$ achieve comparable performance across evaluation metrics, whereas $\mathcal{L}_{{\rm CFM}\text{-}{\bf x}_1}$ objective leads to consistently inferior results. The corresponding training curves are presented in Fig.\,\ref{fig:pesq_sisdr1}, showing that the $\mathcal{L}_{{\rm CFM}\text{-}{\bf x}_1\text{-EDM}}$ model converges in roughly half the time required by the other two. These results suggest that the EDM reweighting stabilizes $\mathbf{x}_1$ training.


\vspace{-0.5cm}
\subsection{\mbox{Performance with additional PESQ and SI-SDR losses}}
We further investigate the effect of incorporating PESQ and SI-SDR losses. As shown in Table\,\ref{tab:pesq_sisdr}, introducing $\mathcal{L}_{\rm PESQ}$ substantially improves PESQ scores. However, this comes at the cost of degraded performance on other metrics, particularly SI-SDR and WER. This observation aligns with the findings of \cite{de2024pesqetarian}, which report that PESQ-oriented optimization may overfit the PESQ metric to the detriment of others.

When combining $\mathcal{L}_{\rm PESQ}$ and $\mathcal{L}_{\rm SI\text{-}SDR}$, the models achieve a better trade‑off across evaluation metrics, as summarized in Table\,\ref{tab:pesq_sisdr}. Under these combined objectives, models trained with different flow matching formulations exhibit comparable overall performance. However, as shown in Fig.\,\ref{fig:pesq_sisdr2}, their training dynamics differ markedly: models optimized with $\mathcal{L}_{{\rm CFM}\text{-}{\bf v}}$ and $\mathcal{L}_{{\rm CFM}\text{-}{\bf x}_1\text{-EDM}}$ converge faster and more stably, whereas the plain $\mathcal{L}_{{\rm CFM}\text{-}{\bf x}_1}$ objective suffers from slower convergence and instability. Moreover, a comparison between Fig.\,\ref{fig:pesq_sisdr1} and Fig.\,\ref{fig:pesq_sisdr2} shows that adding PESQ and SI‑SDR losses largely accelerates convergence and leads to higher PESQ scores.
These results highlight that while PESQ-based optimization alone may bias training, combining perceptual and signal-based criteria leads to more balanced performance and improved training behavior.

\vspace{-0.2cm}
\subsection{Comparison to Baselines}
Based on our findings in Table\,\ref{tab:diffobj} and Table\,\ref{tab:pesq_sisdr}, we recommend using FM with the preconditioned $\mathbf{x}_1$ prediction objective ($\mathcal{L}_{{\rm CFM}\text{-}{\bf x}_1\text{-EDM}}$) combined with PESQ and SI-SDR losses. This configuration provides both rapid convergence and balanced performance. We compare this recommended model against state-of-the-art generative baselines: SGMSE+, a score-matching method; StoRM, its cascaded variant; and SBVE, a Schrödinger bridge model that uses a PESQ loss.

The results are presented in Table\,\ref{tab:baselines}. The upper part shows that our FM model, even without auxiliary losses, achieves the highest PESQ score and the lowest WER among all compared generative models while requiring the fewest sampling steps. The lower part highlights that while SBVE attains the best PESQ score with its specialized loss, it does so at the expense of a substantially reduced SI-SDR. In contrast, our proposed method achieves a more balanced performance across all metrics, confirming its effectiveness and efficiency. These findings strongly support our recommended training strategy.

\vspace{-0.1cm}
\section{Conclusion}
This paper investigated a range of training objectives for flow matching in speech enhancement and analyzed their effects on both training dynamics and final performance. We additionally incorporated perceptual and signal-based loss functions, which not only accelerated convergence but also improved the perceptual quality of the enhanced speech. Although our framework primarily relies on intrusive metric losses for optimization, post-training alignment strategies, such as those explored in \cite{li2025aligning,zhang2025multi}, could be further applied to better match non-intrusive metrics like DNSMOS. Such extensions hold promise for narrowing the gap between objective scores and human perceptual judgments.

\clearpage
\bibliographystyle{IEEEbib}
\bibliography{cited}
\end{document}